\documentclass[letterpaper]{JHEP}%
\usepackage{epsfig}
\usepackage{amsmath}
\usepackage{graphicx}
\usepackage{amsfonts}
\usepackage{amssymb}%
\title{Solving Skyrmions}
\author{Luc Marleau\\D\'{e}partement de physique, g\'{e}nie physique et d'optique, Universit\'{e}
Laval, Qu\'{e}bec QC Canada, G1K 7P4}
\date{March, 2004\\
March, 2004}
\abstract{We find exact solutions for Skyrmions for the Skyrme-like models. Constructing
first the recursion formulae at small and large distance behavior, we proceed
by implementing these constraints to a chosen parametrization of the
solutions. The procedure is applied to the spherically symmetric hedgehog
solution and to topological number $N>1$ solutions based on rational maps.
}
\keywords{Skyrmions, Solitons Monopoles and Instantons , Chiral lagrangians}
\begin{document}

\section{Introduction}

As a non-linear theory of pions, the Skyrme model \cite{Skyrme61} provides an
approximate description of hadronic physics in the low-energy limit. In this
theory, the nucleon emerges as a non-perturbative solution of the field
equations, or more precisely as a topological soliton. This model is also seen
as a prototype which might be applicable in various physical contexts where
one could expect soliton solutions to occur (e.g. condensed matter (baby
Skyrmions), wrapped branes \cite{Marleau02}, ...). More recently, this picture
has regained attention since it could provide an explanation for the newly
discovered hadronic states \cite{newstates03t, newstates03e}.

The original Skyrme lagrangian is a naive extension of the non-linear sigma
model consisting of a fourth-order field derivatives term. This is nonetheless
sufficient to stabilize the soliton against scale transformations and to reach
at least a 30\% accuracy with respect to physical observables. In order to
incorporate effects due to higher-spin mesons and improve the fit on most
observables a number of alternative Skyrme-like models which preserved the
form of the original lagrangian while extending it to higher orders has been
proposed and analyzed \cite{Marleau89-10, Marleau90-13, Jackson91,
Gustafsson94}. Unfortunately, all those models are limited in that they do not
admit exact analytic solutions. Indeed, very few analytic soliton solutions
are known namely the one-dimensional sine-Gordon equation , the KdV equation,
the instanton and some other special cases. This is because a necessary
condition for a soliton-like solution to exist makes the search for an
analytic solution rather complex and its discovery accidental.

In the absence of exact analytical solutions, the only alternative to
numerical treatment is the use of aptly chosen analytical forms which provide
sometimes a reasonable approximation but which may not reproduce the correct
behavior in the limits $r\rightarrow0,\infty$. Apart from greatly simplifying
calculations of physical quantities, a great deal of information can be
extracted from such analytic forms. For instance, symmetries and general
behavior of the solution are much easier to analyze, and the characteristics
of each Skyrme-like model also become more explicit. An analytic form would
also prove useful in the stability analysis of the soliton, both classical and
quantum, and in the calculation of multi-Skyrmion interactions. For example,
one can analyze the quantum behavior of the Skyrme model soliton based on a
family of trial functions, taking into account breathing motions and
spin-isospin rotations or use these solutions to examine the two Skyrmion
interactions. For a more accurate analysis, one has to resort to numerical
computations which can be time consuming (e.g. the full numerical solutions
for lower topological (or baryon number) $N\leq17$ of Houghton et al
\cite{Houghton98}).

The purpose of this work is to find multi-skyrmion solutions analytically
assuming rational maps. For the sake of simplicity, we consider a class of
models which is at most of order six in derivative of the fields. Floratos and
Piette \cite{Floratos01} have indeed shown that for $N\leq5$, the solutions
have the same symmetries as the pure Skyrme model which are well represented
by the rational map ansatz. The general character of the calculation also
allows to extract an exact solution for the $N=1$ hedgehog solution or for
convenience in some cases, simplified approximations of the solutions for any
$N$. We proceed in three steps: First, we write the series expansion of the
solution near $r=0$ and find the recursion relation for the coefficient of the
series (Section III). Second, an analogous analysis is performed in the limit
$r\rightarrow\infty$. Finally, we propose a parametrization for the full
solution and use the former recursion relations to set the parameters (Section
IV). The calculations are somewhat intricate but once the relations are found
they can be easily implemented as a computer algorithm both in numerical or in
symbolic calculations. The method described here is also versatile in the
sense that one could propose a different or more appropriate parametrization
and still use the same recursion relations for $r\rightarrow0,\infty$ to fix
the parameters. We shall conclude with comments on the advantages and
limitations of the procedure and on various ways of improving the convergence
of the series (Section VI) where the chiral angle shows an abrupt behavior
(e.g. larger $N$).

\section{Skyrme model}

The Skyrme model \cite{Skyrme61} is defined by the lagrangian:
\begin{equation}
\mathcal{L}=-\frac{F_{\pi}^{2}}{16}\operatorname*{Tr}L_{\mu}L^{\mu}+\frac
{1}{32e^{2}}\operatorname*{Tr}f_{\mu\nu}f^{\mu\nu}%
\end{equation}
where $L_{\mu}=U^{\dagger}\partial_{\mu}U$ and $f_{\mu\nu}\equiv\lbrack
L_{\mu},L_{\nu}]$ with $U=U(x)$ is an element of the $SU(2)$ group. The first
term, $\mathcal{L}_{1},$ corresponds to the lagrangian of the non-linear
$\sigma$-model. This becomes obvious when one substitutes the degrees of
freedom in $U$ by $\sigma$- and $\mathbf{\pi}$-fields using $U=\frac{2}%
{F_{\pi}}(\sigma+i\mathbf{\tau}\cdot\mathbf{\pi})$ where $\mathbf{\tau}$ are
the three Pauli matrices. The second term, $\mathcal{L}_{2},$ contains four
derivatives of the pion field and can account for nucleon-nucleon interactions
via pion exchange. It was first introduced by Skyrme in order to prevent
solitonic solutions arising in the non-linear $\sigma$-model from shrinking to
zero and thus allowing for stable topological solitons.

The coefficients $F_{\pi}$ and $e$ \ are respectively the pion decay constant
(186 MeV) and a dimensionless coupling often called the Skyrme parameter.
Here, we shall use more appropriate units and rescale the lagrangian according
to
\begin{equation}
\mathcal{L}=\left(  -\frac{1}{2}\operatorname*{Tr}L_{\mu}L^{\mu}\right)
+\frac{1}{2}\left(  \frac{1}{16}\operatorname*{Tr}f_{\mu\nu}f^{\mu\nu}\right)
\label{skyrmelag}%
\end{equation}
With this normalization, lengths can be understood as units of $\frac
{2\sqrt{2}}{eF_{\pi}}$ while energy or mass as units of $\frac{F_{\pi}}%
{2\sqrt{2}e}.$

The lowest energy soliton was found by Skyrme himself and it takes the form%
\begin{equation}
U(\mathbf{r})=\exp\left[  i\boldsymbol{\tau}\cdot\widehat{\mathbf{r}%
}F(r)\right]  \label{hedgehog}%
\end{equation}
where $F(r)$ is called the chiral angle or profile function of the solution,
and $\widehat{\mathbf{r}}$ is radial unit vector. Its spherical symmetry and
the hairy configuration of the spin and isospin pointing out at infinity has
earned it the name of hedgehog solution but more technically, such field
configuration constitutes a map from physical space $R^{3}$ onto the group
manifold $SU(2)$ and is assumed to go to the trivial vacuum for asymptotically
large distances. The latter constraint allows imposing that $U(r\rightarrow
\infty)\rightarrow1$ from which one may derive the existence of a topological
invariant associated with the mapping. The originality of Skyrme's idea was to
identify this invariant, i.e. the winding number $N$,
\[
N=(\operatorname{factor})\int d^{3}rB_{0}\ \ \ \mathrm{with}\ \ \ B_{\mu
}=\epsilon_{\mu\nu\rho\sigma}L^{\nu}L^{\rho}L^{\sigma}%
\]
where $B_{0}$ is the topological charge density, with the baryon number.

For the static hedgehog configuration (\ref{hedgehog}), the energy density is
the sum of
\begin{equation}
\mathcal{E}_{1}=-\frac{1}{2}\operatorname*{Tr}L_{i}L^{i}=[2a+b]
\end{equation}%
\begin{equation}
\mathcal{E}_{2}=-\frac{1}{16}\operatorname*{Tr}f_{ij}f^{ij}=a[a+2b].
\end{equation}
with $a\equiv\frac{\sin^{2}F}{r^{2}}\ \ \mathrm{and}\ \ b\equiv F^{\prime2}.$
Notice that although the second term in (\ref{skyrmelag}) is quartic in the
derivatives of the pion field, the corresponding energy $\mathcal{E}_{2}$
remains only quadratic in $F^{^{\prime}}$.

Despite its relative simplicity, the Skyrme model has deep implications and is
rather successful in describing low-energy hadron physics. Yet, it cannot be
considered seriously as potential candidate for the full low-energy effective
theory of QCD. For example, there is no compelling reason (other than
simplicity) and certainly, no physical grounds to exclude higher-order
derivatives in the pion field from the effective lagrangian. On the contrary,
large $N_{c}$ analysis suggests that the bosonization of QCD should involve an
infinite number of mesons, which implies that in the decoupling limit (or
large mass limit) for higher spin mesons that it leads to an all-orders
lagrangian for pions. Several attempts to construct a more realistic effective
lagrangian were made either by adding vector mesons to the Skyrme picture or
higher-order derivatives terms to the lagrangian \cite{Marleau90-13}.

In the latter approach, one adds a sixth-order term involving the baryonic
density $B_{\mu}$ \cite{Jackson91},
\[
\mathcal{L}_{J}=c_{J}\mathrm{Tr}\ [B^{\mu}B_{\mu}]=3a^{2}b\ \ \ \mathrm{with}%
\ \ \ B_{\mu}=\epsilon_{\mu\nu\rho\sigma}L^{\nu}L^{\rho}L^{\sigma}%
\]
where $c_{J}$ is a constant. The term $\mathcal{L}_{3}=\frac{1}{32}%
\operatorname*{Tr}f_{\mu\nu}f^{\nu\lambda}f_{\lambda}^{\ \ \mu}$ proposed in
\cite{Marleau90-13} leads to an identical contribution to the static energy
density,.%
\begin{equation}
\mathcal{E}_{3}=3a^{2}b
\end{equation}
Allowing for all-orders in derivatives of the pion fields, one can demonstrate
that models can be constructed in terms of $\mathcal{L}_{1},$ $\mathcal{L}%
_{2}$ and $\mathcal{L}_{3}$ alone and that in a rather simple class of these
models \cite{Marleau90-13}, the static energy density is a combination of the
three invariants, $\mathcal{E}_{1},\mathcal{E}_{2}$ and $\mathcal{E}_{3},$
which remains quadratic in $F^{^{\prime}}$ and thus tractable. Moreover the
full lagrangian in this case can be easily written using a generating function
\cite{Marleau01}.

The purpose of the present work is to find multi-skyrmion solutions
analytically and, for the sake of simplicity, we shall only consider a
subclass of these models, those that can be written as a linear combination of
$\mathcal{L}_{1},$ $\mathcal{L}_{2}$ and $\mathcal{L}_{3}$ i.e. the most
general sixth-order lagrangian. The static energy density for the $N=1$
hedgehog ansatz in such a generalized model is simply given by%
\[
\mathcal{E}=\sum_{m=1}^{3}h_{m}\mathcal{E}_{m}=3\chi(a)+(b-a)\chi^{\prime}(a)
\]
where $h_{m}$ are coefficients and $\chi(x)=\sum_{m=1}^{3}h_{m}x^{m}$.
Integrating over volume leads to the mass of the soliton%
\[
M_{S}=4\pi\int_{0}^{\infty}r^{2}dr[3\chi(a)+(b-a)\chi^{\prime}(a)]
\]
Using the same notation, the chiral equation becomes:%

\[
0=\chi^{\prime}(a)\left[  F^{\prime\prime}+2\frac{F^{\prime}}{r}-2\frac{\sin
F\cos F}{r^{2}}\right]  +a\chi^{\prime\prime}(a)\left[  -2\frac{F^{\prime}}%
{r}+F^{\prime2}\frac{\cos F}{\sin F}+\frac{\sin F\cos F}{r^{2}}\right]  .
\]
with $a\equiv\frac{\sin^{2}F}{r^{2}}.$ The Skyrme lagrangian corresponds to
the case $\chi(a)=\chi_{S}(a)\equiv a+\frac{1}{2}a^{2}$.

\section{$N>1$ Skyrmions and rational maps}

For $N>1$ Skyrmions, we shall assume that they are conveniently described by
appropriate rational conformal transformations on the hedgehog solution as
suggested by Houghton et al. \cite{Houghton98}. This assertion has been
verified numerically up to $N=5$ by Floratos and Piette \cite{Floratos01}
where they find that the solutions exhibit the same symmetries as the pure
Skyrme model.

The rational map ansatz lies on a naive connection between rational maps which
are maps from $S^{2}\mapsto S^{2}$ and Skyrmions which are maps from $%
\mathbb{R}
^{3}\mapsto S^{3}.$ It is then possible to identify the domain $S^{2}$ with
concentric spheres in $%
\mathbb{R}
^{3}$, and the target $S^{2}$ with spheres of latitude on $S^{3}.$ Amazingly,
these are accurate to a few percent with respect to lowest energy solutions
obtained through lengthy numerical calculations. Ioannidou et al.
\cite{Ioannidou99} later generalized the ansatz in the context of $SU(N)$
Skyrme models using harmonic maps from $S^{2}$ into $CP^{N-1}.$

Rational maps are usually described in terms of the complex coordinate $z=$
$\tan(\theta/2)e^{i\varphi}$ which correspond to stereographic projections on
conventional polar coordinates $\theta\ $and $\varphi.$ The point $z$
identifies a unit vector
\[
\widehat{\mathbf{n}}_{z}=\frac{1}{1+|z|^{2}}(z+\bar{z},z-\bar{z},1-|z|^{2}).
\]
A rational map is a conformal map defined by
\[
R(z)=\frac{p(z)}{q(z)}%
\]
where one of the polynomials $p(z)\ $and $q(z)$ is at least of degree $N.$ One
can then associate a unit vector to the rational map $R(z)\ $as follows:
\[
\widehat{\mathbf{n}}_{R}=\frac{1}{1+|R|^{2}}(R+\bar{R},R-\bar{R},1-|R|^{2}).
\]

In the context of the Skyrme model, the rational map ansatz propose a solution
of the form
\[
U(r,z)=\exp(i\ \widehat{\mathbf{n}}_{R}\cdot\mathbf{\tau}F(r))
\]
which reduces to the hedghog ansatz (\ref{hedgehog}) when $N=1.$ With boundary
conditions $F(0)=\pi$ and $F(\infty)=0$, the baryon number turns out to be
$N$, the degree of $R(z)$. Using this ansatz, the static energy density for
this class of models takes the form
\[
\mathcal{E}=\sum_{m=1}^{\infty}h_{m}a_{N}^{m-1}[3a_{N}+m(b_{N}-a_{N})]
\]
where now $b_{N}=F^{\prime2}=b$ and $a_{N}=a\rho(z)$ with
\[
\rho(z)=\left(  \frac{1+|z|^{2}}{1+|R|^{2}}\left\vert \frac{dR}{dz}\right\vert
\right)  ^{2}.
\]
A remarkable advantage of the rational maps lies in the fact that the
separation of the angular and radial dependence of the solution is preserved.
Integrating over to obtain the mass of the soliton, we have%

\[
M_{S}=\int_{0}^{\infty}r^{2}dr\int d\Omega\sum_{m=1}^{\infty}h_{m}\left(
a\rho\right)  ^{m-1}[(3-m)a\rho+mb]
\]
where the integration over solid angle reads
\[
\int d\Omega\rightarrow\int_{-\infty}^{\infty}\frac{2i\ dzd\bar{z}}%
{(1+|z|^{2})^{2}}%
\]
and the factor $2i\ dzd\bar{z}/(1+|z|^{2})^{2}$ is equivalent to the usual
area element on a 2-sphere $\sin\theta d\theta d\varphi.$

Defining
\[
I_{m}^{N}=\int\frac{d\Omega}{4\pi}\rho^{m}=\frac{1}{4\pi}\int\left(
\frac{1+|z|^{2}}{1+|R|^{2}}\left\vert \frac{dR}{dz}\right\vert \right)
^{2m}\frac{2i\ dzd\bar{z}}{(1+|z|^{2})^{2}}.
\]
and
\begin{equation}
\chi_{1}(a)=\sum_{m=1}^{\infty}\alpha_{m}a^{m}\qquad\chi_{2}(a)=\sum
_{m=1}^{\infty}\beta_{m}a^{m} \label{alphabeta}%
\end{equation}
where $\alpha_{m}=h_{m}I_{m}^{N},$ $\beta_{m}=h_{m}I_{m-1}^{N},$ we can write
the expression for multi-Skyrmion masses%
\begin{equation}
M_{S}=4\pi\int_{0}^{\infty}r^{2}dr\left(  3\chi_{1}(a)-a\chi_{1}^{\prime
}(a)+b\chi_{2}^{\prime}(a)\right)  \label{masseq}%
\end{equation}
Some of the angular integrations are trivial for rational maps: $I_{0}^{N}=1,$
$I_{1}^{N}=N,$ $I_{m}^{1}=1$. Of course, the solutions depend on the model,
i.e. the weight of each terms in $\chi_{1}$ and $\chi_{2}$, or more precisely
on the value of the each coefficient $h_{m}$. This in turn determines which
rational maps $R,$ and values of angular integrations $I_{m}^{N},$ minimizes
the mass.

From now on, we shall restrict our analysis to models of order six in
derivatives of the pion field by setting coefficients $h_{m}=0$ for $m>3$.
These models correspond to the most general case where the choice of rational
maps affects a single non-trivial angular integration $I_{2}^{N}$ and the
expression for the mass simplifies according to
\begin{align*}
\chi_{1}(a)  &  =h_{1}Na+h_{2}I_{2}^{N}a^{2}\\
\chi_{2}(a)  &  =h_{1}a+h_{2}Na^{2}+h_{3}I_{2}^{N}a^{3}.
\end{align*}
Using the Derrick arguments, a stable soliton exist for $N=1$ only if $h_{3}$
is positive. This term prevents the Skyrmion from shrinking to zero size
against a scale transformation. In this case, to find the minimal energy
configuration at fixed $N$ (fixed degree for $R(z)$) one proceeds as follows:
(i) minimize $I_{2}^{N}$ as a function of the coefficients of polynomials
$p(z)\ $and $q(z)$ and (ii) find the profile function $F(r)$ which minimizes
the energy. Note that since $I_{2}^{N}$ is positive and can go up to infinity
for $N>1$, negative values $h_{2}$ are physically excluded, as minimization
would lead to an infinitely large negative soliton mass. One then concludes
that rational map configurations and symmetries which minimizes the mass for a
general sixth-order model (positive $h_{1},h_{2}$ and $h_{3}$) are the same as
for the Skyrme model since one must simply minimize $I_{2}^{N}$ in both cases.

The chiral equation becomes accordingly:%
\begin{align}
0  &  =h_{1}\left(  F^{\prime\prime}+2\frac{F^{\prime}}{r}\right)  +2N\left(
-h_{1}\left[  \frac{\sin F\cos F}{r^{2}}\right]  +h_{2}\left(  \frac{\sin
^{2}F}{r^{2}}\right)  \left[  F^{\prime\prime}+F^{\prime2}\frac{\cos F}{\sin
F}\right]  \right) \nonumber\\
&  +I_{2}^{N}\left(  -2h_{2}\left(  \frac{\sin^{2}F}{r^{2}}\right)  \left[
\frac{\sin F\cos F}{r^{2}}\right]  +3h_{3}\left(  \frac{\sin^{2}F}{r^{2}%
}\right)  ^{2}\left[  F^{\prime\prime}+2\frac{F^{\prime}}{r}\right]  \right)
\label{chiralfeq}%
\end{align}

Our goal is to obtain an analytic expression for the solution of the chiral
equation. For this purpose, we write the chiral equation in a form that is
more convenient using $\phi=\cos F:$%

\begin{align}
0  &  =-2r^{2}\chi_{1}^{\prime}\left(  1-\phi^{2}\right)  ^{2}\phi+\chi
_{1}^{\prime\prime}\left(  1-\phi^{2}\right)  ^{3}\phi\\
&  +\chi_{2}^{\prime}\left(  -\phi^{\prime\prime}\left(  1-\phi^{2}\right)
r^{2}-\left(  \phi^{\prime}\right)  ^{2}\phi r^{2}-2r\left(  1-\phi
^{2}\right)  \phi^{\prime}\right)  +\chi_{2}^{\prime\prime}\left(  2r\left(
1-\phi^{2}\right)  \phi^{\prime}+r^{2}\phi\left(  \phi^{\prime}\right)
^{2}\right)  \left(  1-\phi^{2}\right)  \label{chiraleqfi}%
\end{align}
where $\chi_{i}\equiv\chi_{i}\left(  \frac{\left(  1-\phi^{2}\right)  }{r^{2}%
}\right)  .$

We would like to represent its solution $\phi(r)$ in the form of a series. Its
explicit form is a matter of choice as long as it reproduces the solution
adequately over all values of $r$ which requires the convergence of the
series. Otherwise, what choice of series is best would have to be judged on a
number of criteria which have to be met at least minimally: (1) rapid
convergence, i.e. fewer terms are necessary to reach numerical precision with
respect to the exact solution, (2) mathematical tractability e.g. recursion
formula for the coefficients of the series can be written and (3) other
aesthetical criteria, e.g. a form of series leading to an energy density and
other physically quantity that can be integrated analytically.

Solutions can be constructed simply by adding terms to approximate solutions
proposed in the past
\cite{Balachadran82,Atiyah-Manton89,Neto91,Mignaco92,Marleau94,Kopeliovich01,Ponciano01}%
. However, a quick analysis reveals that some of these candidates must be
discarded since they do not behave properly at small or large distances. In
this work, we propose a solution which implements these constraints and a
procedure to allow for the convergence of the series. However, the first step
of our analysis will be independent of a specific choice of series
representation for the solution. It consists in finding the small
($r\rightarrow0$) and large ($r\rightarrow\infty$) distance behavior of the
solution as a series expansion in powers $r$ and $1/r$ respectively. This is
computed using the chiral equation. Once the small and large $r$ expansion are
known explicitly, the information will serve to propose an appropriate form
for the series and ultimately to calculate the coefficients of the series.

\section{Small distance behavior\label{small}}

First, let us write $\phi$ as a power expansion in $r$ in the limit
$r\rightarrow0.$ The chiral equation (\ref{chiraleqfi}) is symmetric with
respect to the change $r\rightarrow-r$, $\phi(r)\rightarrow\phi(-r)=\phi(r)$,
which simplifies the power expansion since only even powers of $r$
contribute:
\begin{equation}
\phi(r)=\sum_{n=0}^{\infty}a_{n}r^{2n} \label{smallrsln}%
\end{equation}
with boundary condition $a_{0}=-1$ since $\phi(0)=-1.$

For convenience, let us write explicitly the functions $\chi_{1}$ and
$\chi_{2}$ in the chiral equation (\ref{chiraleqfi})
\begin{align}
0  &  =r^{2}\phi\sum_{n=0}^{\infty}n(n-3)\alpha_{n}a^{n+1}+\phi\left(
\phi^{\prime}\right)  ^{2}\sum_{n=0}^{\infty}n(n-2)\beta_{n}a^{n-1}%
\label{chiraleqfi2}\\
&  +2r\phi^{\prime}\sum_{n=0}^{\infty}n(n-2)\beta_{n}a^{n}-r^{2}\phi
^{\prime\prime}\sum_{n=0}^{\infty}n\beta_{n}a^{n}.\nonumber
\end{align}
Substituting $\phi$ by the series (\ref{smallrsln}) and $\phi^{\prime}%
,\phi^{\prime\prime},\phi\left(  \phi^{\prime}\right)  ^{2}$ and $a^{n}$ by
the appropriate expressions
\[%
\begin{tabular}
[c]{ccc}%
$\phi\left(  \phi^{\prime}\right)  ^{2}=\sum_{m=0}^{\infty}b_{m}r^{2m}$ &
$\qquad$ & $a^{n}=\sum_{k=0}^{\infty}c_{n,k}r^{2k}$%
\end{tabular}
\ \
\]
in the last equation, we get
\begin{align*}
0  &  =\sum_{n=1}^{3}\sum_{k=0}^{\infty}\sum_{l=0}^{\infty}\left[
r^{2}n(n-3)\alpha_{n}\left(  c_{n+1,k}r^{2k}a_{l}r^{2l}\right)  +n(n-2)\beta
_{n}\left(  c_{n-1,k}r^{2k}b_{l}r^{2l}\right)  \right. \\
&  \left.  +2rn(n-2)\beta_{n}\left(  2c_{n,k}r^{2k}la_{l}r^{2l-1}\right)
-r^{2}n\beta_{n}\left(  2c_{n,k}r^{2k}l(2l-1)a_{l}r^{2l-2}\right)  \right]  .
\end{align*}

The coefficients $b_{k}$ and $c_{n,k}$ are given respectively by%
\[%
\begin{array}
[c]{ll}%
b_{1}=4a_{0}a_{1}^{2} & \\
b_{k}=\bar{b}_{k}+\Delta b_{k} & \quad\text{for }k>1
\end{array}
\]%
\[%
\begin{array}
[c]{ll}%
c_{0,0}=1 & \\
c_{0,k}=0 & \quad\text{for }k>0\\
c_{n,0}=(-2a_{0}a_{1})^{n} & \quad\text{for }n>0\\
c_{n,k}=\bar{c}_{n,k}+\Delta c_{n,k} & \quad\text{for }n,k>0
\end{array}
\]
with
\begin{align*}
\bar{b}_{k}  &  =4a_{0}\sum_{q=2}^{k-1}q(k+1-q)a_{q}a_{k+1-q}+4\sum
_{m=1}^{k-1}a_{m}\left(  \sum_{q=1}^{k-m}q(k+1-m-q)a_{q}a_{k+1-m-q}\right) \\
\Delta b_{k}  &  =8ka_{0}a_{1}a_{k}\\
\bar{c}_{n,k}  &  =\frac{1}{2ka_{0}a_{1}}\left(  \sum_{q=1}^{k-1}%
(q(n+1)-k)2a_{0}a_{q+1}c_{n,k-q}+\sum_{q=1}^{k}(q(n+1)-k)\sum_{r=1}^{q}%
a_{r}a_{q+1-r}c_{n,k-q}\right) \\
\Delta c_{n,k}  &  =-2^{n}na_{0}a_{1}^{n-1}a_{k+1}%
\end{align*}
where for later convenience, we have isolated the contribution proportional to
the coefficients $a_{k}$ with highest $k$ in the terms $\Delta b_{k}$ and
$\Delta c_{n,k}$ respectively.

The equation holds for arbitrary $r$, which implies that it holds for any
arbitrary power of $r$. Isolating the term in $r^{2(k-1)}$ we get an
expression of the form
\begin{equation}
0=A_{k-1}a_{k}+B_{k-1}.\label{recursionsmall}%
\end{equation}
where $A_{k-1}$ and $B_{k-1}$ are respectively
\[
A_{k-1}=\sum_{n=1}^{3}2n\left(  2a_{1}\right)  ^{n}\left(  -\alpha_{n}\left(
n-3\right)  \left(  n+1\right)  +\beta_{n}\left(  n^{2}-2n-2-2k^{2}+k\right)
\right)  ,
\]%
\begin{align*}
B_{k-1} &  =\sum_{n=1}^{3}n\sum_{m=2}^{k-1}(n-3)\alpha_{n}\left(
a_{m}c_{n+1,k-m-1}+a_{0}\bar{c}_{n+1,k-1}+a_{1}c_{n+1,k-2}\right)  \\
&  +\beta_{n}((n-2)\left(  b_{m}c_{n-1,k-m}+\bar{b}_{k}c_{n-1,0}+b_{1}\bar
{c}_{n-1,k-1}\right)  \\
&  \quad\quad\quad+2m\left(  2n-3-2m\right)  a_{m}c_{n,k-m}+2(2n-5)a_{1}%
\bar{c}_{n,k-1}).
\end{align*}
From inspection we see that both $A_{k-1}$ and $B_{k-1}$ depend on lower-index
coefficients $a_{m}$ (i.e. $m<k)$ so we can write (\ref{recursionsmall}) in
the form of a recursion formula for the coefficients $a_{k}$ for $k>1,$%

\[
a_{k}=-\frac{B_{k-1}}{A_{k-1}}\text{\qquad with \qquad}a_{0}=-1.
\]

The formula requires a single input parameter $a_{1}$ (which must be
positive), aside from the boundary condition $a_{0}=-1$. $a_{1}$ is related to
the slope of the profile function at $r=0$ as follows
\[
F(r\rightarrow0)=\pi-\sqrt{2a_{1}}r+\mathcal{O}(r^{3})\qquad\text{or\qquad
}\phi(r\rightarrow0)=-1+a_{1}r^{2}+\mathcal{O}(r^{4}).
\]

It is also easy to verify that $a_{k}$ depend on all $\alpha_{n}$'s and
$\beta_{n}$'s whose values are prescribed by both the model and the
topological number of the solution (see (\ref{alphabeta})). This has an
interesting implication: The small $r$ behavior of the solution is
characterized by all the terms in the lagrangian and not only the one with
highest-order derivative in the pion field as one might have suspected.

\section{Large distance behavior}

Now let us examine $\phi$ as a power expansion in $1/r$ in the limit
$r\rightarrow\infty$ \ Again since the chiral equation (\ref{chiraleqfi}) is
symmetric with respect to the change $r\rightarrow-r$, $\phi(r)\rightarrow
\phi(-r)=\phi(r)$, the power expansion only contains even powers of $1/r$.
Accordingly, we write
\begin{equation}
\phi=\sum_{m=0}^{\infty}\hat{a}_{m}r^{-2m}. \label{largersln}%
\end{equation}
Allowing for the boundary condition $\phi(\infty)=1,$ we set $\hat{a}_{0}=1$.
It is also easy to verify that
\[
\hat{a}_{1}=0\text{\quad and \quad}\hat{a}_{3}=0.
\]
According to (\ref{largersln}), we write the following expression as power
series
\[%
\begin{tabular}
[c]{ccc}%
$\phi\left(  \phi^{\prime}\right)  ^{2}=\sum_{m=0}^{\infty}\widehat{b}%
_{m}r^{-2m}$ & $\qquad$ & $a^{n}=\sum_{m=0}^{\infty}\widehat{c}_{n,m}r^{-2m}$%
\end{tabular}
\ \
\]
where the coefficients $\widehat{b}_{k}$ and $\widehat{c}_{n,k}$ are given
respectively by%

\[%
\begin{array}
[c]{ll}%
\widehat{b}_{5}=16\hat{a}_{0}\hat{a}_{2}^{2} & \\
\widehat{b}_{k}=\widehat{\bar{b}}_{k}+\Delta\widehat{b}_{k} & \quad\text{for
}k>5
\end{array}
\]%
\[%
\begin{array}
[c]{ll}%
\widehat{c}_{n,0}=0 & \\
\widehat{c}_{n,k}=0 & \quad\text{for }k<3n\\
\widehat{c}_{n,3n}=\left(  -\right)  ^{n} & \\
\widehat{c}_{n,k}=\widehat{\bar{c}}_{n,k}+\Delta\widehat{c}_{n,k} &
\quad\text{for }k>3n
\end{array}
\]
with%
\begin{align*}
\widehat{\bar{b}}_{k}  &  =4\hat{a}_{0}\sum_{l=3}^{k-4}l(k-1-l)\hat{a}_{l}%
\hat{a}_{k-1-l}+4\sum_{m=2}^{k-1}\hat{a}_{m}\sum_{l=2}^{k-3-m}l(k-1-m-l)\hat
{a}_{l}\hat{a}_{k-1-m-l}\\
\Delta\widehat{b}_{k}  &  =16(k-3)\hat{a}_{0}\hat{a}_{2}\hat{a}_{k-3}\\
\widehat{\bar{c}}_{n,k}  &  =-n\left(  -2\hat{a}_{0}\hat{a}_{2}\right)
^{n-1}\sum_{q=2}^{k-3n}\hat{a}_{q}\hat{a}_{k-3n+2-q}-\frac{1}{2\hat{a}_{0}%
\hat{a}_{2}\left(  k-3n\right)  }\sum_{q=1}^{k-3n-1}(q(n+1)-k+3n)\widehat
{c}_{1,q+3}\widehat{c}_{n,k-q}\\
\Delta\widehat{c}_{n,k}  &  =-2\hat{a}_{0}n\left(  -2\hat{a}_{0}\hat{a}%
_{2}\right)  ^{n-1}\hat{a}_{k-3n+2}.
\end{align*}
Here the contributions proportional to the coefficients $a_{k}$ with highest
$k$ are written explicitly in $\Delta\widehat{b}_{k}$ and $\Delta\widehat
{c}_{n,k}$ respectively.

Substituting $\phi$ by the series (\ref{largersln}) $\phi^{\prime}%
,\phi^{\prime\prime},\phi\left(  \phi^{\prime}\right)  ^{2}$ and $a$ by the
above expressions we arrive at
\begin{align}
0  &  =\sum_{n=0}^{\infty}\sum_{k=0}^{\infty}\sum_{m=0}^{k}\left[
n(n-3)\alpha_{n}\hat{a}_{m}\widehat{c}_{n+1,k-m}r^{-2k+2}+n(n-2)\beta
_{n}\widehat{b}_{m}\widehat{c}_{n-1,k-m}r^{-2k}\right. \\
&  \left.  -2n(n-2)\beta_{n}2m\hat{a}_{m}\widehat{c}_{n,k-m}r^{-2k}-\sum
_{n=0}^{\infty}n\beta_{n}2m(2m+1)\hat{a}_{m}\widehat{c}_{n,k-m}r^{-2k}\right]
.
\end{align}
Again, since the equation must hold for arbitrary $r,$ it implies that it
holds for an arbitrary power of $r$. Therefore isolating the term in
$r^{-2k+3}$ leads to the relation
\begin{equation}
0=\widehat{A}_{k-1}\hat{a}_{k}+\widehat{B}_{k-1} \label{recursionlarge}%
\end{equation}
with%
\[
\widehat{A}_{k-1}=4\hat{a}_{2}\left(  -4\alpha_{1}+\beta_{1}\left(
2k^{2}-5k+6\right)  \right)
\]%
\begin{align*}
\widehat{B}_{k-1}  &  =-2\alpha_{1}\left(  \hat{a}_{0}\widehat{\bar{c}%
}_{2,k+4}+\sum_{m=1}^{k-2}\left(  \hat{a}_{m}\widehat{c}_{2,k-m+4}\right)
\right) \\
&  -\beta_{1}\left(  12\hat{a}_{2}\widehat{\bar{c}}_{1,k+1}+\sum_{m=0}%
^{k+2}\left(  \widehat{b}_{m}\widehat{c}_{0,k-m+3}\right)  +\sum_{m=3}%
^{k-1}2m\left(  2m-1\right)  \hat{a}_{m}\widehat{c}_{1,k-m+3}\right) \\
&  +\sum_{n=2}^{\infty}\sum_{m=0}^{k+3}\left(  n(n-3)\alpha_{n}\hat{a}%
_{m}\widehat{c}_{n+1,k-m+4}+n(n-2)\beta_{n}\widehat{b}_{m}\widehat
{c}_{n-1,k-m+3}\right. \\
&  \left.  -2n\left(  2n-3+2m\right)  m\beta_{n}\hat{a}_{m}\widehat
{c}_{n,k-m+3}\right)  .
\end{align*}
After inspection, we find that the right hand side of equation
(\ref{recursionlarge}) is trivially zero for $k<3$ and but otherwise contains
only terms in $a_{m}$ with $m\leq k.$ Furthermore, $\widehat{A}_{k-1}$ and
$\widehat{B}_{k-1}$ only depend on lower-index coefficients $a_{m}$ ($m\leq
k-1$).$\ $This allows writing a recursion formula for the coefficients
$\hat{a}_{k}$ for $k\geq3,$%
\[
\hat{a}_{k}=-\frac{\widehat{B}_{k-1}}{\widehat{A}_{k-1}}.
\]
Recalling that $\hat{a}_{0}=1$, $\hat{a}_{1}=0$ and $\hat{a}_{3}=0$, the only
remaining unknown parameter is $\hat{a}_{2}$ which must be negative$.$ This
latter parameter fixes the dominant contribution to the profile function in
the large $r$ limit,
\[
F(r\rightarrow\infty)=\frac{\sqrt{-2\hat{a}_{2}}}{r^{2}}+\mathcal{O}(\frac
{1}{r^{6}})\qquad\text{or\qquad}\phi(r\rightarrow\infty)=1+\frac{\hat{a}_{2}%
}{r^{4}}+\frac{\hat{a}_{4}}{r^{8}}+...
\]
Moreover, the first coefficients $\hat{a}_{k}$ in the large $r$ expansion
depend only on the lowest-$n$ $\alpha_{n}$'s and $\beta_{n}$'s. This means
that the large-$r$ behavior is not very sensitive to the model and to the
topological sector of the solution. In other words, higher-order derivative
terms and the topological sector begin to contribute only when the
sub-dominant terms in $r^{-2k}$ become important.

\section{Solutions for Skymions}

We have found recursion relations for the coefficients of a series expansion
in powers $r$ and $1/r$ for small ($r\rightarrow0$) and large ($r\rightarrow
\infty$) distances respectively. We may now proceed to construct a full
solution. The explicit form of the series is to a certain extent a matter of
choice. Of course, it should represent the solution adequately which means
that convergence is a prerequisite and that small ($r\rightarrow0$) and large
($r\rightarrow\infty$) distance behaviors must be reproduced. Other criteria
may also be suggested such as: (1) faster convergence, i.e. fewer terms are
necessary to reach numerical precision with respect to the exact solution, (2)
mathematical tractability e.g. recursion formula for the coefficients of the
series can be written, (3) the energy density can be integrated analytically, etc...

We propose here a solution and a procedure similar to that in \cite{Marleau94}
but which applies to $\phi=\cos F$ instead of $F$. Taking in account the small
($r\rightarrow0$) and large ($r\rightarrow\infty$) distance behavior, we can
write recursion relations for the coefficients of the series. We shall see
that this particular ansatz also allows in principle, to calculate the mass of
the soliton analytically which was not possible in \cite{Marleau94}. The
solution has the form%

\begin{equation}
\phi_{r_{0}}(r)=\phi(\frac{r}{r_{0}})=1+\sum_{m=0}^{\infty}\frac{\left(
r^{2}\right)  ^{m}\left(  r_{0}^{2}\right)  ^{m+2}}{\left(  r^{2}+r_{0}%
^{2}\right)  ^{2m+3}}\left(  c_{2m}(r^{2}+r_{0}^{2})+c_{2m+1}r_{0}^{2}\right)
\label{solutionfi}%
\end{equation}

The small distance expansion is given by%

\begin{align*}
\phi_{r_{0}}(r) &  =1+\left(  c_{0}+c_{1}\right)  +\sum_{q=0}^{\infty}\left(
\frac{r^{2}}{r_{0}^{2}}\right)  ^{2q+2}(\frac{(q+3)(q+2)\left(  -1\right)
^{q+1}}{2}\left(  c_{0}+c_{1}\right)  \\
&  \quad+\sum_{k=0}^{q}\frac{(2q-k+2)!\left(  -1\right)  ^{k}}{k!(2q-2k+4)!}%
((2q-2k+4)(2q-2k+3)c_{2q-2k}\\
&  \quad\quad\quad\quad+(2q-k+4)(2q-k+3)\left(  c_{2q-2k+2}+c_{2q-2k+3}%
\right)  )\\
&  =\left(  1+c_{0}+c_{1}\right)  +\left(  -2c_{0}-3c_{1}+c_{2}+c_{3}\right)
\left(  \frac{r^{2}}{r_{0}^{2}}\right)  ^{2}+\left(  3c_{0}+6c_{1}%
-4c_{2}-5c_{3}+c_{4}+c_{5}\right)  \left(  \frac{r^{2}}{r_{0}^{2}}\right)
^{4}+...
\end{align*}

Matching the coefficients at small distance with those of
(\ref{recursionsmall}), we get%
\[
a_{0}=1+c_{0}+c_{1}=-1
\]
and for $k\geq1$%
\begin{align*}
\left(  r_{0}^{2}\right)  ^{k}a_{k} &  =\frac{(k+2)(k+1)\left(  -1\right)
}{2}^{k}(a_{0}-1)+c_{2k-2}+c_{2k}+c_{2k+1}+\sum_{r=1}^{k-1}\frac
{(2k-r)!\left(  -1\right)  ^{r}}{r!(2k-2r+2)!}\\
&  \quad\cdot\left(  (2k-2r+2)(2k-2r+1)c_{2k-2r-2}+(2k-r+2)(2k-r+1)\left(
c_{2k-2r}+c_{2k-2r+1}\right)  \right)  .
\end{align*}
We invert the relation to get the odd-index coefficients and obtain%
\begin{equation}
c_{1}=a_{0}-1-c_{0}=-2-c_{0}\label{c1}%
\end{equation}
and for $k\geq1$%
\begin{align}
c_{2k+1} &  =\left(  r_{0}^{2}\right)  ^{k}a_{k}-c_{2k}-c_{2k-2}%
-\frac{(k+2)(k+1)\left(  -1\right)  }{2}^{k}(a_{0}-1)\\
&  -\sum_{r=1}^{k-1}\frac{(2k-r)!\left(  -1\right)  ^{r}}{r!(2k-2r+2)!}%
((2k-2r+2)(2k-2r+1)c_{2k-2r-2}\label{codd}\\
&  \quad\quad+(2k-r+2)(2k-r+1)\left(  c_{2k-2r}+c_{2k-2r+1}\right)  ).
\end{align}

On the other hand, the large distance behavior reads%
\begin{align*}
\phi_{r_{0}}(r  & \rightarrow\infty)=1+c_{0}\left(  \frac{r^{2}}{r_{0}^{2}%
}\right)  ^{-2}+\sum_{q=0}^{\infty}\left(  \frac{r^{2}}{r_{0}^{2}}\right)
^{-q-3}\\
& \cdot\left(  c_{2q+2}+\sum_{k=0}^{q}\frac{(2q-k+2)!\left(  -1\right)  ^{k}%
}{\left(  k+1\right)  !(2q-2k+2)!}\left(  2\left(  -q+k-1\right)
c_{2q-2k}+\left(  k+1\right)  c_{2q-2k+1}\right)  \right)  .
\end{align*}
Again by inspection%
\[
-2c_{0}+c_{1}+c_{2}=0\text{\quad or\quad}c_{2}=2c_{0}-c_{1}=3c_{0}+2
\]
and matching the coefficients at large distance, we get the even-index
coefficients of the solution which then reads
\begin{equation}
c_{2k}=\hat{a}_{k+2}\left(  r_{0}^{2}\right)  ^{-k-2}-\sum_{r=0}^{k-1}%
\frac{(2k-r)!\left(  -1\right)  ^{r}}{\left(  r+1\right)  !(2k-2r)!}\left(
2\left(  -k+r\right)  c_{2k-2r-2}+\left(  r+1\right)  c_{2k-2r-1}\right)
.\label{ceven}%
\end{equation}

Inserting (\ref{c1}), (\ref{codd}) and (\ref{ceven}) in (\ref{solutionfi}), we
end up with a solution $\phi_{r_{0}}$ which is now completely determined by
three parameters, $a_{1}$, $\hat{a}_{2}$ and $r_{0}$. The first two parameters
$a_{1}$, $\hat{a}_{2}$ depend on the behavior near $r\rightarrow0$ and
$r\rightarrow\infty$ respectively whereas $r_{0}$ can be interpreted as an
intermediate scale. The strategy adopted here to reach the solution lies upon
the fact that for a given $k,$ the even-index coefficients $c_{2k}$ in
(\ref{ceven}) depend on lower-index coefficients $c_{m}$ with $m<2k$ and
$\hat{a}_{k+2}$ which is fixed by the $(k+2)^{\text{th}}$ coefficient of the
large $r$ expansion or ultimately $\hat{a}_{2}$. On the other hand, the
odd-index coefficients $c_{2k+1}$ in (\ref{codd}) requires $c_{m}$ with
$m<2k+1$ and $a_{k}$ which is determined by the behavior of solution at small
$r$ or the parameter $a_{1}$. This has two effects: First, the procedure
refines the solution by alternatively the matching of the odd- and even-index
coefficients, which requires higher and higher derivatives of the solution at
$r\rightarrow0$ and $r\rightarrow\infty$ respectively (\ref{solutionfi}).
Secondly, since it relies on matching the coefficients $a_{k}$ and $\hat
{a}_{k+2},$ the accuracy of the series and convergence must improve near the
end points as $k$ increases.

It remains that the parameters $a_{1},$ $\hat{a}_{2}$ and $r_{0}$ are still
unknown at this point of the procedure. Several approaches to find these
parameters are possible. One such procedure consists in introducing three more
constraints on $\phi_{r_{0}}$. For example, we can use the chiral equation
(\ref{chiraleqfi}) at three intermediate points to set the values of $a_{1},$
$\hat{a}_{2}$ and $r_{0}$. We could also choose to impose a continuity
condition at a given $r=r_{0}$ assuming that the series (\ref{smallrsln}) and
(\ref{largersln}) apply to $r\leq r_{0}$ and $r\geq r_{0}$ respectively. This
procedure would lead to a solution which has the virtue of being completely
determined by the equation of motion, but they do not guarantee that the set
of parameters would be the best to render the exact solution. Since these
methods would require numerical calculations at some point anyhow, we adopt a
more practical approach which consists of finding numerically the set of
values for $a_{1},$ $\hat{a}_{2}$ and $r_{0}$ which minimize the mass of the soliton.

In principle, the series (\ref{solutionfi}) contains an infinite number of
terms which allow in the end to reach the exact the solution. For
computational reason the series is truncated which means that, the number of
terms in the series will affect not only the precision of the series but also
the values of the parameters $a_{1},$ $\hat{a}_{2}$ and $r_{0}$ that minimize
the mass of the soliton. The larger number of terms in the series, the closer
we get to the exact solution and values of the parameters $a_{1},$ $\hat
{a}_{2}$ and $r_{0}$.

We could attempt to compute the mass of the soliton analytically. Starting
from (\ref{masseq}) we have%
\[
M_{S}=4\pi\int_{0}^{1}2u^{\frac{3}{2}}\left(  1-u\right)  ^{\frac{1}{2}%
}du\left(  h_{1}\left(  2Na+b\right)  +h_{2}a\left(  I_{2}^{N}a+2Nb\right)
+3h_{3}I_{2}^{N}a^{2}b\right)
\]
where we changed variable $r$ to $r\rightarrow u=\frac{r^{2}}{r^{2}+r_{0}^{2}%
}$ or $r=r_{0}\sqrt{\frac{u}{1-u}}$. Substituting $\phi$ by our solution%

\begin{equation}
\phi_{r_{0}}(u)=1+\sum_{k=0}^{K}u^{k}\left(  1-u\right)  ^{k+2}\left(
c_{2k}+c_{2k+1}\left(  1-u\right)  \right)  \label{solutionapprox}%
\end{equation}
where $K$ is finite but sufficiently large for accuracy, and assuming that we
can expand the integrand in powers $u$\ and $\left(  1-u\right)  ,$ it becomes
possible to integrate the expression analytically since all integrals can be
cast in the form of%
\[
M_{ab}=\int_{0}^{1}u^{a-1}(1-u)^{b-1}du=\frac{\Gamma(a)\Gamma(b)}{\Gamma
(a+b)}\qquad\text{ for Re}(a),\text{Re}(b)>0.
\]
Unfortunately, even for the simplest case, the Skyrme model, the calculation
becomes inefficient and quite impractical as $K$ increases. So we resort to
numerical integration which in this case is much faster and proceed to
minimize the mass in terms of the parameters $a_{1},$ $\hat{a}_{2}$ and
$r_{0}$ . As an example, we show the result of the first few values of $K$ for
the Skyrme model,. for $N=1,$ which requires $h_{1}=h_{2},$ $h_{3}=0$ and
$I_{1}^{1}=I_{2}^{1}=1$
\[%
\begin{tabular}
[c]{cccccc}\hline\hline
& $K=1$ & $K=2$ & $K=3$ & $K=4$ & Num.\\\hline
Mass & $1.23268$ & $1.23174$ & $1.23151$ & $1.23151$ & $1.23145$\\
$a_{1}$ & $2.17741$ & $2.05754$ & $2.02667$ & $1.99411$ & $2.01508$\\
$\hat{a}_{2}$ & $-2.39428$ & $-1.62650$ & $-1.22111$ & $-3.15231$ &
$-2.33204$\\
$r_{0}$ & $1.05398$ & $1.12095$ & $1.18501$ & $0.939329$ & $-$%
\end{tabular}
\ \ \ \
\]
The exact mass is obtained within an accuracy of $0.005\%$ for $K$ as low as
$3.$ Note however that the convergence of $a_{1}$ and $\hat{a}_{2}$ towards
their exact values is not as efficient.

\section{Advantages and limitations of the approach}

An extensive comparison of the approximate solution (\ref{solutionapprox})
with respect to the exact numerical solution suggests that the reliability of
the approach depends largely on the model considered and topological sector.
The recursion formulae for the $c_{k}$ coefficients ensure a perfect agreement
in both limits $r\rightarrow0$ and $r\rightarrow\infty$. The agreement is
still preserved when the solution remains smooth for all values of $r$, as in
the case of low $N$ soliton in the Skyrme Model. Nevertheless, discrepancies
between analytical and numerical results begin to appear when we consider
sixth-order models or when the topological number $N$ increases. The solutions
in these cases are characterized by a sharp behavior of $F$ near $F(r)=\pi/2$
(or $\phi$ near $\phi(r)=0)$.

In order to understand the origin of these discrepancies, it is instructive to
look at the quantity $a=r^{-2}\sin^{2}F$ for small, large and intermediate
$r$. For the intermediate region where $\sin^{2}F$ reaches its maximum, the
energy density is found to be the largest and $F$ is almost linear. On the
other hand, $F^{\prime\prime}$ and of course $\cos F$ are relatively small.
Therefore, which term dominates the chiral equation (\ref{chiralfeq}) in that
region depends on the relative weight of the coefficients $h_{1},$
$2Nh_{1},2Nh_{2},$ $I_{2}^{N}h_{2}$ and $I_{2}^{N}h_{3}$ and on the highest
power of $a$. As $N$ increases, $I_{2}^{N}$ increases approximately as $N^{2}$
and the term proportional to $h_{3}I_{2}^{N}$ is expected to dominate if
$h_{2}$ and $h_{3}$ are of the order same of magnitude. This implies that most
of the variation of $F$ should occur in the intermediate region since the
dominant term is proportional to $a^{2}$ which is suppressed outside this
region. This leads to a configuration of the energy density which is localized
on a shell of decreasing thickness as $N$ increases. Unfortunately, the
recursion formulae for the $c_{k}$ coefficients only apply for the end regions
$r\rightarrow0$ and $r\rightarrow\infty$ which do not contribute much to the
soliton mass. \ So we can expect the approximate solution
(\ref{solutionapprox}) to lose its accuracy when $N$ gets larger. In other
word, although accuracy is expected to increase in the $r\rightarrow0$ and
$r\rightarrow\infty$ region, it does not improve in the intermediate region
where most of the energy is concentrated.

Various ways of improving the analytical solution may be considered. For
instance, several other trial functions have been used in conjunction with the
small and large $r$ recursion formulae, including polynomials series and
Pad\'{e} approximants, but we found no general form that would eliminate the
aforementioned discrepancies. In fact, there is no reason a priori why any
trial function should evade the conclusion of the last paragraph. A better
improvement would fix the behavior of the profile angle and describe
accurately the energy density in the intermediate region where its
contribution is more important for any $N.$ This could proceed through the
construction of a recursion formula at an intermediate point and eventually
its implementation into a series but, it implies rather complex intricate
calculations and has not been attempted yet.

Even though imperfect, we stress again that this approach is still probably
the best alternative to complete numerical treatment, and that it can prove
very useful whenever an analytical form of the solution is required, as we
have already discussed regarding stability analysis. Because of the general
character of the calculation, it easy find the exact solution for the $N=1$
hedgehog solution for the pure Skyrme model. It is also possible --- if one
chooses not use the full solution --- to extract approximate solutions for any
$N$ since the first few terms of the series represent fairly good
approximations$.$ Moreover, this work also provide the first construction of
the recursion formulae for the series expansion in the $r\rightarrow0$ and
$r\rightarrow\infty$ limits which may be helpful in many calculations.

This research was supported by the Natural Science and Engineering Research
Council of Canada.


\begin{thebibliography}{99}                                                                                               %


\bibitem {Skyrme61}T.H.R. Skyrme, Proc.~R.~Soc.~London.~A2603 (1961) 127.

\bibitem {Marleau02}F. Leblond and L. Marleau, JHEP 0205 $\left(  2002\right)
,$035.

\bibitem {newstates03t}D. Diakonov, V. Petrov and M. V. Polyakov, Z. Phys. A
359, $\left(  1997\right)  $ 305; H. Weigel, Eur. Phys. J. A 2 $\left(
1998\right)  ,$391; M. Prasza lowicz, Phys. Lett. B 575 $\left(  2003\right)
$ 234.

\bibitem {newstates03e}T. Nakano et al. [LEPS Collaboration], Phys. Rev. Lett.
91 (2003) 012002. V. V. Barmin et al., Phys. Atom. Nucl. 66 (2003) 1715 [Yad.
Fiz. 66 (2003) 1763]; S. Stepanyan et al. , hep-ex/0307018. J. Barth et al. ,
hep-ex/0307083; V. Kubarovsky and S. Stepanyan and CLAS Collaboration,
hep-ex/0307088; A. E. Asratyan, A. G. Dolgolenko and M. A. Kubantsev,
hep-ex/0309042. V. Kubarovsky et al., , hep-ex/0311046; A. Airapetian et al.,
arXiv:hep-ex/0312044; S. Chekanov, http://www.desy.de/f/seminar/Chekanov.pdf.

\bibitem {Marleau89-10}L. Marleau, Phys.~Lett.~B235 (1990) 141 .

\bibitem {Marleau90-13}S. Dub\'{e} and L. Marleau, Phys.~Rev.~D41, (1990) 1606.

\bibitem {Jackson91}A.D. Jackson, C. Weiss and A. Wirzba, Nucl.~Phys.~A529
(1991) 741 .

\bibitem {Gustafsson94}K. Gustafsson and D.O. Riska, Nucl. Phys. A571 (1994) 645.

\bibitem {Houghton98}C.J. Houghton, N.S. Manton and P.M. Sutcliffe, Nucl.
Phys. B510, (1998) 507.

\bibitem {Floratos01}I. Floratos and B. Piette,. Phys.Rev.D64 (2001) 045009.

\bibitem {Marleau01}L. Marleau and J.F. Rivard,. Phys.Rev.D63 (2001) 036007.

\bibitem {Ioannidou99}T.Ioannidou, B.Piette and W.J.Zakrzewski, J.Math.Phys.
40 (1999) 6353.

\bibitem {Balachadran82}A.P. Balachandran, V.P. Nair, S.G. Rajeev and A.
Stern,. Phys.Rev.Lett.49 (1982) 1124; Erratum-ibid.50 (1983) 1630;
Phys.Rev..D27 $\left(  1983\right)  $ 1153; Erratum-ibid.D27 $\left(
1983\right)  $ 2772.

\bibitem {Atiyah-Manton89}M.F. Atiyah and N.S. Manton (Cambridge U.),
Phys.Lett.B222 (1989) 438.

\bibitem {Neto91}J. Ananias Neto, R. Mendez Galain and E. Ferreira,
J.Math.Phys.32 (1991)1949.

\bibitem {Mignaco92}J.A. Mignaco and S. Wulck, J.Phys.G18 $\left(
1992\right)  $1309; Erratum-ibid.G18 $\left(  1992\right)  $ 2061.

\bibitem {Marleau94}B. Dion and L. Marleau, Phys.Rev.D49 $\left(  1994\right)
$ 5526.

\bibitem {Kopeliovich01}V B. Kopeliovich, JETP Lett.73 $\left(  2001\right)  $
587; Pisma Zh.Eksp.Teor.Fiz.73 $\left(  2001\right)  $ 667.

\bibitem {Ponciano01}J.A. Ponciano, L.N. Epele, H. Fanchiotti and C.A. Garcia
Canal, Phys.Rev.C64 $\left(  2001\right)  $ 045205.
\end{thebibliography}
\end{document}